# Determination of the magnetization profile of Co/Mg periodic multilayers by magneto-optic Kerr effect and X-ray magnetic resonant reflectivity

**P Jonnard[1], K Le Guen[1], J-M André[1], R Delaunay[1], N Mahne[2], A Giglia[2], S Nannarone[2,3], A Verna[2], Z-S Wang[4], J-T Zhu[4], S-K Zhou[4]**

[1] Laboratoire de Chimie Physique – Matière et Rayonnement, Université Pierre et Marie Curie, CNRS UMR 7614, 11 rue Pierre et Marie Curie, F-75231 Paris Cedex 05, France
[2] Istituto Officina dei Materiali IOM-CNR Laboratorio TASC, SS 14 km 163,5, I-34149 Basovizza, Trieste, Italy
[3] Dipartimento di Ingegneria dei Materiali e dell Ambiente, Universita di Modena e Reggio Emilia, Via Vignolese 905, I-41100 Modena, Italy
[4] Institute of Precision Optical Engineering, Department of Physics, Tongji University, Shanghai 200092, P. R. China

E-mail: philippe.jonnard@upmc.fr

**Abstract.** The resonant magnetic reflectivity of Co/Mg multilayers around the Co $L_{2,3}$ absorption edge is simulated then measured on a specifically designed sample. The dichroic signal is obtained when making the difference between the two reflectivities measured with the magnetic field applied in two opposite directions parallel to the sample surface. The simulations show that the existence of magnetic dead layers at the interfaces between the Co and Mg layers leads to an important increase of the dichroic signal measured in the vicinity of the third Bragg peak that otherwise should be negligible. The measurements are in agreement with the model introducing 0.25 nm thick dead layers. This is attributed to the Co atoms in contact with the Mg layers and thus we conclude that the Co-Mg interfaces are abrupt from the magnetic point of view.

## 1. Introduction

The Co/Mg periodic multilayers are promising for optics applications in the EUV range between 20 and 30 nm [1,2]. Indeed, we have measured an experimental reflectivity of 0.43 has been reported and up to 51% for Co/Mg-based multilayers [3,4]. Due to the high quality of the Co/Mg interfaces [1], this value is about 75% of the simulated value for a perfect stack, *i.e.* with perfect interfaces without roughness or interdiffusion. The quality of the interfaces was checked from the physico-chemical point of view by x-ray emission spectroscopy (XES) and nuclear magnetic resonance (NMR) spectroscopy. Both were in agreement and did not indicate any noticeable interdiffusion between the Co and Mg layers. However, given the magnetic character of Co, it is interesting to probe the internal structure of the system through a magneto-optic effect, particularly to check if the interfaces are also abrupt from the magnetic point of view.

Verna et al. [5] have demonstrated the possibility to determine the magnetic profiles at manganite surfaces with a monolayer resolution by using resonant magnetic reflectivity. Using the same kind of methodology and taking advantage of the magnetic character of the Co layers, we expect to be able to obtain a precise determination of the magnetization profile within the Co/Mg multilayer. This is made by measuring the reflectivity as a function of the incident photon energy around the Co $L_3$ and $L_2$ edges, i.e. around 778 and 793 eV, and as a function of the Bragg angle, i.e. the grazing angle, for a given circularly polarized radiation and for two opposite magnetizations of the sample.

**2. Experiment simulation**

Simulations of the reflectivity are performed using the Pythonic Programming for Multilayer (PPM) [6]. The simulations are carried out for a previously studied multilayer: [Co (2.55 nm) / Mg (5.45 nm)]$_{\times 30}$ and considering a Si substrate [1]. The reflectivity simulations in $\theta$-$2\theta$ mode are performed for two opposite helicities of the incoming radiation and at two photon energies: 777.6 eV, corresponding to the maximum of x-ray absorption coefficient; 776.2 eV, off the absorption maximum, but where the dichroism is still appreciable. Results for the two energies and for $\theta$ up to 26° are presented in figure 1. In this case we report the average of the reflectivities obtained for right and left circularly polarized light. Bragg peaks are much more pronounced at 776.2 eV. This is because out of the absorption maximum, low absorption case, the radiation penetrates deep within the stack and all the 30 bilayers can diffract the incident radiation. Also fringes due to the finite thickness of the overall multilayer are visible. At 777.6 eV, the light is absorbed in the first layers of the stack and then the Bragg peaks are much more smoothed.

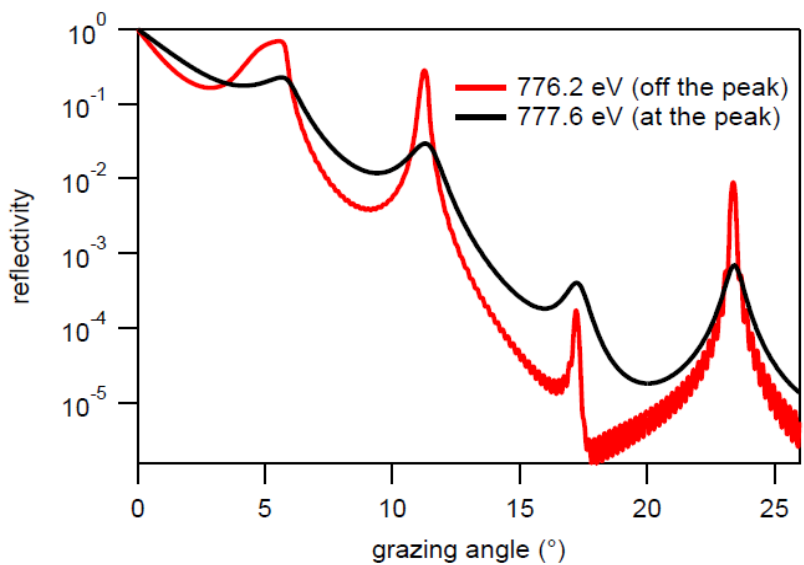

**Figure 1.** Reflectivity simulations of a Co/Mg multilayer at two photon energies corresponding to the maximum of the absorption peak off this peak.

In order to test the sensitivity of the technique on the magnetization profile of the Co layers, we consider three different magnetization profiles of the Co layer as indicated in figure 2. Profile A corresponds to a fully magnetized Co layer. Profile B considers two 0.25 nm thick dead layers at both Co/Mg interfaces, while for profile C the thickness of the dead layers are 0.5 nm thick. The same profile is taken for all the 30 Co layers of the multilayer. The reflectivity simulations are performed as a function of the photon energy around the Co $L_3$ absorption edge and at the fixed grazing angles corresponding to the first three Bragg peaks (n=1, $\theta$=5.72°; n=2 $\theta$=11.50°; n=3 $\theta$=17.39°). The dichroic signal, which is the difference between the reflectivity obtained with the two opposite helicities of the incident radiation, is reported in figure 3 for the three different magnetization profiles of figure 2.

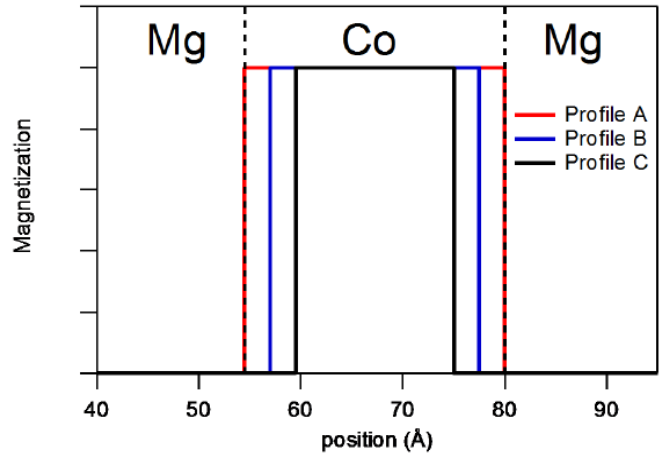

**Figure 2.** Magnetization profiles (in arbitrary units) of the Co layers considered to simulate the magnetic resonant reflectivity of Co/Mg multilayer.

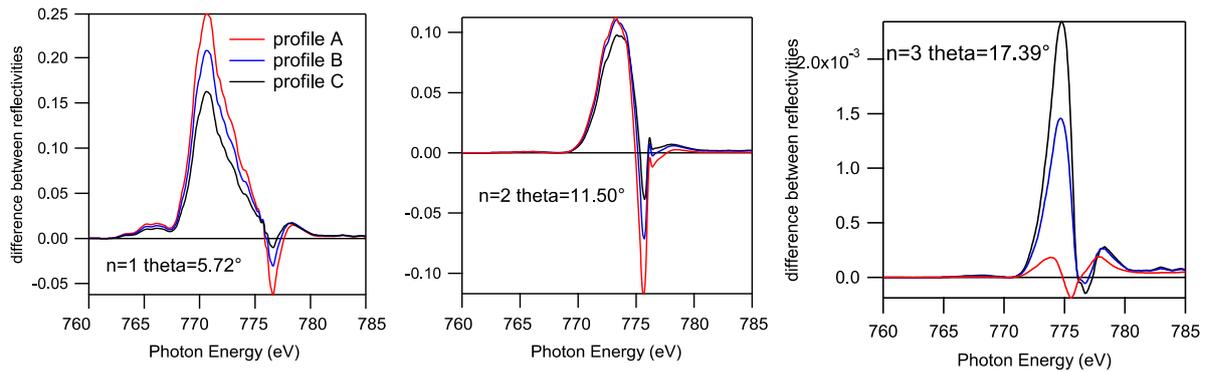

**Figure 3.** Dichroic signal simulated for Co/Mg multilayer around the Co $L_3$ edge, for three different grazing angles corresponding to the first three Bragg angles and for the three magnetization profiles of the Co layers described in the figure 2.

For the Bragg peaks diffracted at the first and second diffraction orders, the dichroic signal is not very sensitive to the magnetization profile. In particular n=2 peak remains almost unchanged, while n=1 peak changes the amplitude but not its sign. A large difference is instead evident for the third Bragg peak. With no dead layer the dichroic signal is weak because in this case, the thickness of the Co layer (2.55 nm) is close to a third of the period thickness (8.0 nm). The existence of a dead layer induces a departure to this ratio and leads to a 10 times increase of dichroic signal.

## 3. Sample design and characterization

Considering the simulations, we design a Co/Mg multilayer which should not present 3$^{rd}$ diffraction order, in fact 3n diffraction orders. Thus if any magnetic dead layer exists at the Co layer interfaces it should give a relatively intense dichroic signal at the 3$^{rd}$ order diffraction peak. So the aimed values of the Co and Mg thicknesses are 2.55 and 5.10 nm respectively, giving a period of 7.65 nm. A multilayer following this design having 30 Co/Mg bilayers was prepared by magnetron sputtering. The deposition was made on a silicon substrate. A 3 nm thick Cr adhesion layer was first deposited onto the substrate. Finally a 3.5 nm thick $B_4C$ capping layer was deposited on top of the stack to prevent its oxidation. After its, preparation the stack was characterized by x-ray reflectivity at 0.154 nm. The reflectivity curve is shown on figure 4. It is observed that the 3$^{rd}$ Bragg peak is almost extinguished and that the 6[th] Bragg peak is suppressed. This demonstrates that the actual layer thicknesses are closed to the designed ones. This is confirmed by the fit of the reflectivity curve. The thicknesses of the Co and Mg layers deduced from the fit, 2.60 and 5.16 nm, very slightly larger than the designed ones, and the interfacial roughnesses are around 0.5 nm. These values represent the *rms* thickness variation of the Co and Mg layers.

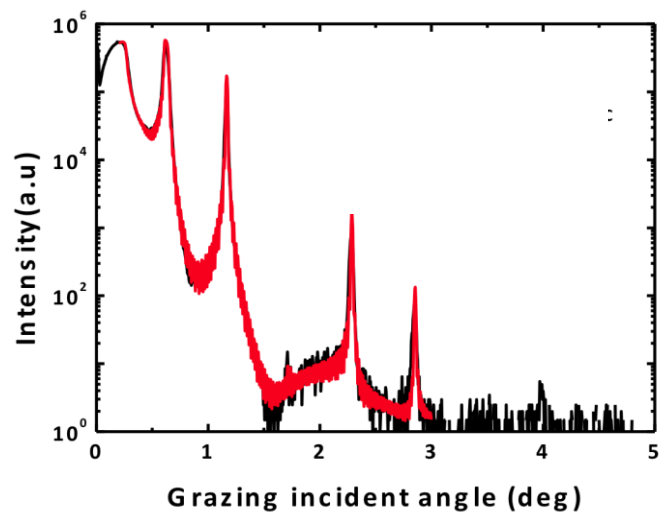

**Figure 4.** Measured (black line) and fitted (red line) hard x-ray reflectivity curves of the prepared Co/Mg multilayer.

Then MOKE (magneto-optic Kerr effect) experiment is made at the wavelength of the He-Ne laser, *i.e.* 632.8 nm. The sample is a 10 mm x 5 mm rectangle. Figure 5 shows the result when the surface of the sample is perpendicular to the applied magnetic field. The signal on the ordinates, is proportional to the rotation of the light polarization. Its relative variation during a cycle is proportional to the magnitude of the magnetization variation after a linear background subtraction. On the range of the hysteresis loop, this relative intensity variation is 1%. The width of the hysteresis loop allows the determination of a coercitive field of about 600 Oe.

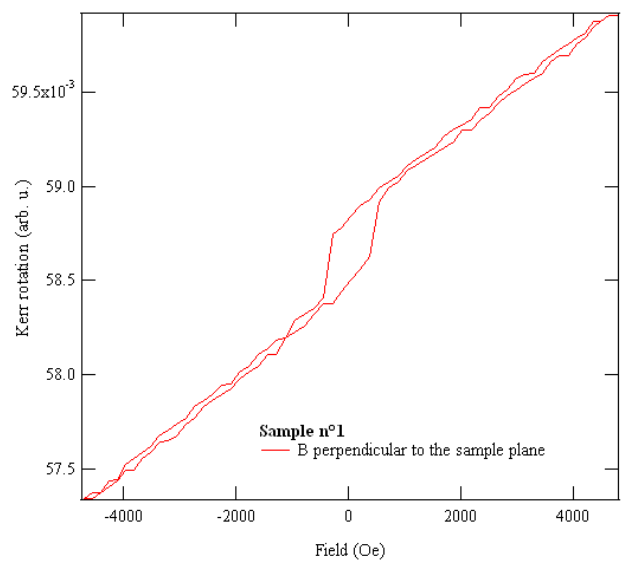

**Figure 5.** MOKE hysteresis curve of the Co/Mg multilayer. The magnetic field is oriented in the direction perpendicular to the sample plane.

Figure 6 (left) shows the MOKE hysteresis when the surface of the sample is parallel to the applied magnetic field and in our setup the sample is fixed horizontally. The coercive field is about 90 Oe. The hysteresis curve is not centered around 0 G because of the remanent field of the electro-magnet (about 40 G). In this case the variation of the magnetization is much larger with respect to the perpendicular configuration: on the range of the hysteresis loop, the relative intensity variation is 6%. So the easy magnetization axis is parallel to the sample surface. Figure 6 (right) shows the result when the surface of the sample is parallel to the applied magnetic field and in our setup the sample is fixed vertically, *i.e.* rotated by 90° with respect to the previous case. The curve is almost the same as in the previous case, that is to say there exists no easy magnetization axis for one of the two sides of the rectangle.

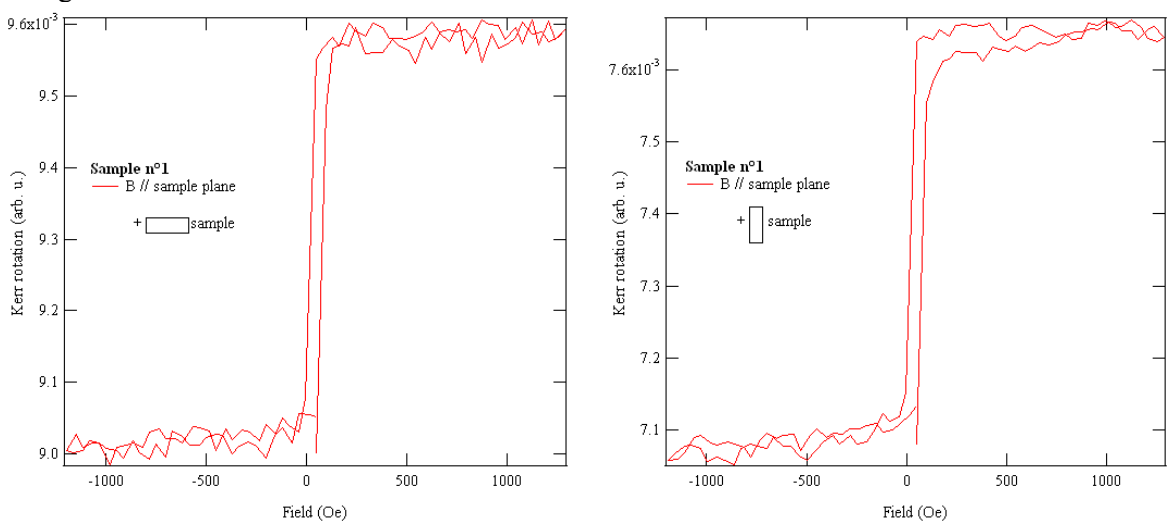

**Figure 6.** MOKE hysteresis curve of the Co/Mg multilayer. The magnetic field is oriented in the direction parallel to the sample plane. Left: horizontal sample; right: vertical sample.

## 4. Results and discussion

The magnetic resonant reflectivity measurements are made by scanning the incident photon energy around the Co $L_3$ and $L_2$ edges and the grazing angle around the five first Bragg peaks. The incident is almost circularly polarized. The measured degree of circular polarization is 0.8. The magnetic field is applied parallel to the surface of the sample, *i.e.* along the easy magnetization axis as determined by MOKE. The intensity of the applied magnetic field is about 400 Oe, sufficient for a complete saturation of the magnetization. To obtain the dichroic signal, the difference between reflectivities measured with opposite magnetizations is calculated. This kind of experiment is equivalent to the simulated one: use two opposite helicites of the radiation and one magnetization of the sample. Experiments were performed at the BEAR beamline of the Elettra synchrotron centre.

First, we show in figure 7, the reflectivity curves and the dichroic signal around the Co $L_{2,3}$ edge at an angle corresponding to the first Bragg peak. The reflectivity curves are presented without applied magnetic field and in the two cases where the magnetic field is applied in opposite directions. The differences between these three cases are small but can be observed because of the anomaly close to the absorption edge and of the high brightness of the synchrotron radiation. The dichroic signal presents a peak followed by a dip at the $L_3$ edge and the same at the $L_2$ edge but with different amplitudes of the peak and dip. The same kind of signal is obtained when making dichroism of the x-ray absorption.

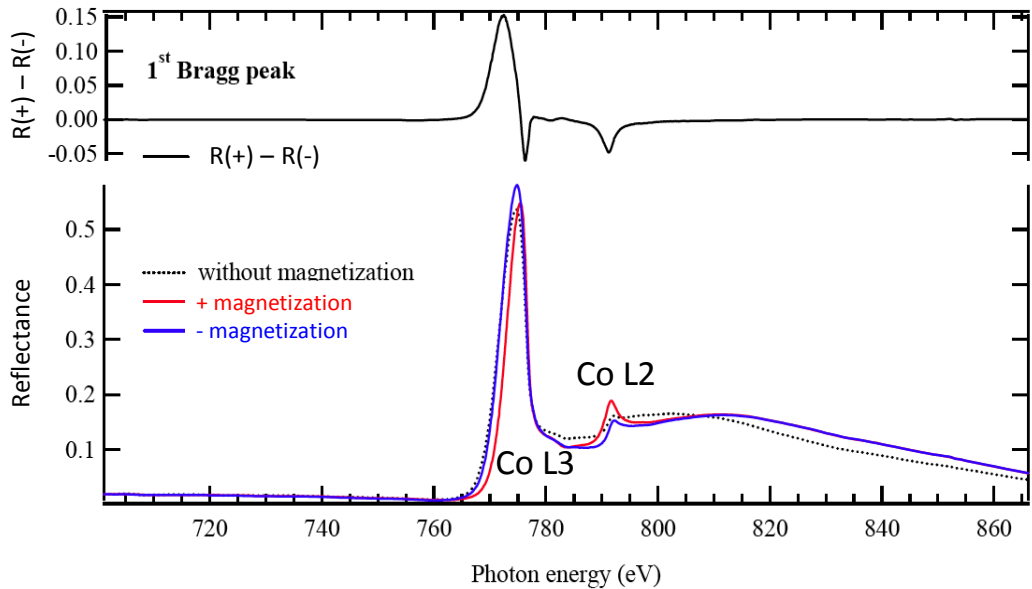

**Figure 7.** Reflectivity curves, with and without applied magnetic field, and dichroic signal around the Co L edges and at an angle corresponding to the first Bragg peak.

Then, we present on figure 8 the 3D maps obtained by varying both the photon energy (only around Co $L_3$ absorption edge) and the grazing angle. For sake of clarity we only present the result obtained around the 1$^{st}$ and 3$^{rd}$ Bragg peaks. Around the 1$^{st}$ Bragg peak the dichroic signal is rather intense and the structures such as the ones observed in figure 7 are recognized. For the 3$^{rd}$ Bragg peak, the curves obtained at an angle corresponding to the 3$^{rd}$ Bragg peak being presented in figure 9, no negative value is present in the dichroic signal whose intensity is quite weak and no dip is present in the spectra as a function of energy for every fixed grazing angle. This indicates the presence of a dead layer at the Co interfaces with Mg. The thickness of this dead layer cannot be comparable with the thickness of a Co layer, because this should imply a breaking of the 2:1 thickness ratio and a strong increase of the 3$^{rd}$ Bragg peak with respect to the non-magnetized case, that is not observed. The amplitude of the signal around the 3rd Bragg peak, peak, $1.6\ 10^{-3}$, is compatible with the simulation with a 0.25 nm thick dead layers, and for which only a very small dip is calculated. Also the maximum of the dichroism in the 1$^{st}$ Bragg peak is close to the one simulated with the 0.25 nm thick dead layer.

Thus considering the closeness of the simulated and deposited stacks, we assume the existence of a 0.25 nm thick magnetic dead layer. This thickness represents about the thickness of one Co atomic plane (0.21 nm). Thus the dead layer only results from the Co interfacial atoms in contact with the Mg atoms of the neighbouring layers. These Co atoms have already been observed by NMR spectroscopy [1], giving a NMR signal in the ratio of number of interfacial Co atoms with respect to number of bulk Co atoms. Then, we conclude that the interfaces within the Mg/Co stacks are abrupt from both magnetic and chemical point of views.

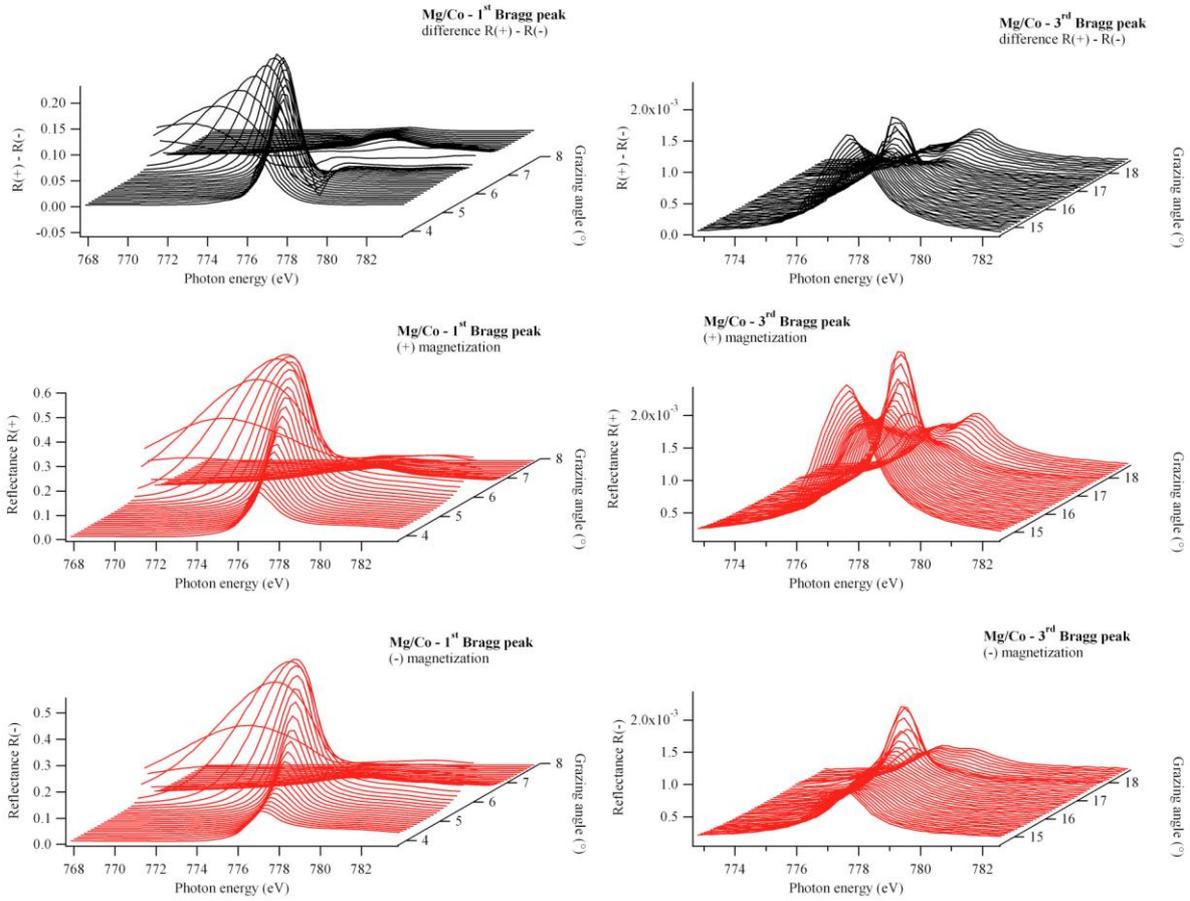

**Figure 8.** Resonant magnetic reflectivity maps of the Co/Mg multilayer, obtained for incident photon energies around the Co $L_3$ edge and grazing angles around the 1st (left) and 3rd (right) Bragg peaks. The reflectivity curves are in red and for the two opposite directions of the applied magnetic field, $R(+)$ and $R(-)$. The dichroic signal is plotted in black.

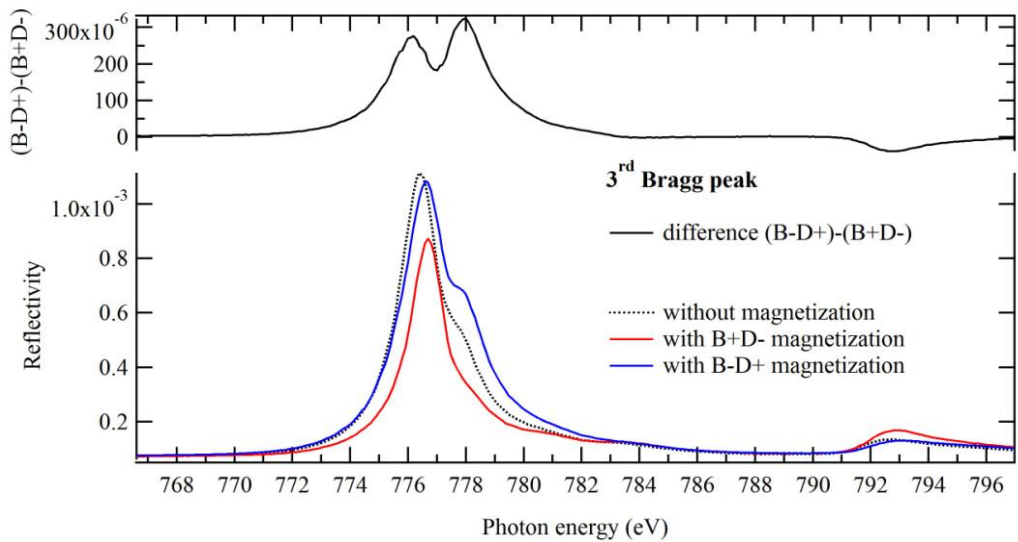

**Figure 9.** Reflectivity curves, with and without applied magnetic field, and dichroic signal around the Co L edges and at an angle corresponding to the thrid Bragg peak.

**Acknowledgments**
Funding from ANR through project COBMUL N°ANR-10-INTB-902-01 is acknowledged. A.V. acknowledges support from the FVG Regional Project SPINOX funded by Legge Regionale 26/2005 and the decreto 2007/LAVFOR/1461. Authors are thankful to George Kourousias and Alessio Curri, Scientific Computing team of the IT group at Elettra synchrotron, for their help in the installation and use of PPM program.